\documentclass{aa}
\usepackage{txfonts}
\usepackage{graphicx}
\newcommand{\magpt}[2]{\mbox{$\rm #1\hspace{-0.25em}\stackrel{m}{.}
      \hspace{-1.0mm}#2$}}                             

\newcommand\teff{$ {\rm T_{eff}}$}

\newcommand\logg{$\log {\rm g}$}
\newcommand\loghe{${\rm \log{\frac{n_{He}}{n_{H}}}}$}

\newcommand{\Msolar}{\mbox{\,$\rm M_{\odot}$}}        
\begin{document}

\title{The hot horizontal-branch stars in $\omega$\,Centauri\thanks{Based on observations
  with the ESO Very Large Telescope at Paranal Observatory, Chile
  (proposal IDs 075.D-0280(A) and 077.D-0021(A))}}
\author{S. Moehler\inst{1}
\and S. Dreizler\inst{2}
\and T. Lanz\inst{3}
\and G. Bono\inst{4,5}
\and A.V. Sweigart\inst{6}
\and A. Calamida\inst{1,5}
\and M. Nonino\inst{7}}
\institute{
European Southern Observatory, Karl-Schwarzschild-Str. 2, D 85748
  Garching, Germany {\sl e-mail: smoehler,acalamid@eso.org}
\and Georg-August-Universit\"at, Institut f\"ur Astrophysik,
Friedrich-Hund-Platz 1, D 37077 G\"ottingen, Germany {\sl e-mail: 
  dreizler@astro.physik.uni-goettingen.de} 
\and Department of Astronomy, University of Maryland, College Park, MD  
20742-2421, USA {\sl e-mail: lanz@astro.umd.edu}
\and Dipartimento di Fisica, Universit\'a di Roma Tor Vergata,
Via della Ricerca Scientifica 1, 00133 Roma, Italy {\sl e-mail:
bono@roma2.infn.it}
\and INAF - Rome Astronomical Observatory, via Frascati 33, 00040 
Monte Porzio Catone, Italy
\and NASA Goddard Space Flight Center, Code 667, Greenbelt, MD 20771,
  USA {\sl e-mail: Allen.V.Sweigart@nasa.gov}
\and INAF - Trieste Astronomical Observatory, via G.B. Tiepolo 11,
40131 Trieste, Italy {\sl e-mail: nonino@ts.astro.it}
}
\date{Received / Accepted }

\abstract {UV observations of some massive globular clusters have
revealed a significant population of stars hotter and fainter than the
hot end of the horizontal branch ({\bf HB}), the so-called blue hook
stars. This feature might be explained either by the late hot flasher
scenario where stars experience the helium flash while on the white
dwarf cooling curve or by the progeny of the helium-enriched
sub-population postulated to exist in some clusters. Previous
spectroscopic analyses of blue hook stars in $\omega$\,Cen and
NGC\,2808 support the late hot flasher scenario, but the stars contain
much less helium than expected and the predicted C and N
enrichment cannot be verified.}
{We compare the observed effective temperatures, surface gravities, helium
  abundances, and carbon line strengths (where detectable) of our
  targets stars with  the predictions of the two
  aforementioned scenarios.}
{Moderately high resolution spectra of hot HB stars in the globular
cluster $\omega$\,Cen were analysed for radial velocity variations,
atmospheric parameters, and abundances using LTE and non-LTE model
atmospheres.}
{We find no evidence of close binaries among our target
  stars. All stars below 30\,000\,K are helium-poor and very similar
  to HB stars observed in that temperature range in other globular
  clusters. In the temperature range 30\,000\,K to 50\,000\,K, we find
  that 28\% of our stars are helium-poor (\loghe $< -1.6$), while 72\%
  have roughly solar or super-solar helium abundance (\loghe $\ge
  -1.5$). We also find that carbon enrichment is strongly
  correlated with helium enrichment, with a maximum carbon
  enrichment of 3\% by mass. }
{A strong carbon enrichment in tandem with helium enrichment is
predicted by the late hot flasher scenario, but not by the
helium-enrichment scenario. We conclude that the helium-rich HB stars
in $\omega$\,Cen cannot be explained solely by the helium-enrichment
scenario invoked to explain the blue main sequence.}
\maketitle

\section{Introduction}
\label{sec:intro}
UV-visual colour-magnitude diagrams of the two very massive globular
clusters, $\omega$\,Cen and NGC\,2808, display a rather puzzling
``hook-like'' feature at the hot end of their extended horizontal
branches with stars lying below the canonical horizontal branch
(Whitney et al. \cite{whro98}; D'Cruz et al. \cite{dcoc00}; Brown et
al. \cite{brsw01}). Similar features were observed in NGC\,2419
(Ripepi et al. \cite{ricl07}), NGC\,6273 (Piotto et
al. \cite{pizo99}), NGC\,6715 (Rosenberg et al. \cite{rore04}), and 
NGC\,6388 (and possibly NGC\,6441, Busso et al. \cite{buca07}).
These blue-hook stars cannot be explained within the framework of
canonical stellar evolution.  Brown et al.\ (\cite{brsw01}) 
proposed a ``flash-mixing'' scenario to explain the blue hook stars.
According to this scenario, stars that lose an unusually large amount
of mass will leave the red giant branch ({\bf RGB}) before the helium
flash and will move quickly to the (helium-core) white dwarf cooling
curve before igniting helium (Castellani \& Castellani \cite{caca93};
D'Cruz et al.\ \cite{dcdo96}; Brown et al. \cite{brsw01}).  However,
the evolution of these ``late hot helium flashers'' differs
dramatically from the evolution of stars that undergo the helium flash
on the RGB. When a star flashes at the tip of the RGB or
shortly thereafter, the large entropy barrier of its strong
hydrogen-burning shell usually prevents the products of helium burning from
being mixed to the surface.  These canonical stars will evolve to the
zero-age horizontal branch ({\bf ZAHB}) without any change in their
hydrogen-rich envelope composition. In contrast, stars that ignite
helium on the white dwarf cooling curve, where the hydrogen-burning
shell is much weaker, will undergo extensive mixing between the
helium- and carbon-rich core and the hydrogen envelope (Sweigart
\cite{swei97}; Brown et al. \cite{brsw01}; Cassisi et
al. \cite{casc03}, \cite{casa09}; Miller Bertolami et
al. \cite{mial08}). Depending on where the helium flash occurs along
the white dwarf cooling curve, the envelope hydrogen will be mixed
either deeply into the core (``deep mixing'') or only within a
convective shell in the outer part of the core (``shallow
mixing''). In the case of deep mixing, virtually all of the envelope
hydrogen is burned, while in shallow mixing some of the envelope
hydrogen remains after the mixing phase (Lanz et al. \cite{labr04}).
One of the most robust predictions of the flash-mixing scenario is an
increase in the surface abundance of carbon to 3\% - 5\% (deep mixing)
or 1\% (shallow mixing) by mass. This increase is set by the carbon
production during the helium flash and is nearly independent of the
stellar parameters.  

For these reasons, the flash convection zone will have a composition
of about 4\% carbon and 96\% helium (plus the minor heavier elements)
as it grows outward through the core toward the hydrogen shell.  Since
there is initially no hydrogen in the core, none of this $3\alpha$
carbon is burned to nitrogen via CNO burning at this stage of the
flash.  This situation changes, however, once the flash convection
zone penetrates into the hydrogen shell, and hydrogen is mixed into
the core.  This hydrogen will burn on the $^{12}$C
in the flash convection zone around the point where the mixing
timescale is comparable to the timescale of the $^{12}$C$+$proton
reaction.  If the number of protons is smaller than or of the
order of the number of $^{12}$C nuclei, the primary outcome of this
hydrogen burning is the production of $^{13}$C.  If, however, there
are of the order of two protons for each $^{12}$C nuclei, then the
$^{12}$C can react to produce some $^{14}$N.  How much $^{14}$N is
produced will depend on the details of the mixing process.  It
appears, however, that the primary outcome of this hydrogen burning is
the production of $^{13}$C with only a much smaller amount of
$^{14}$N.  Thus the predicted surface abundance by mass following
flash mixing should be approximately 96\% helium, 4\% carbon, and
possibly a small amount of nitrogen.  For both deep and shallow
mixing, the blue hook stars should be helium-rich relative to the
canonical extreme HB (EHB) stars.

On the other hand, some authors propose that the blue hook stars are
  the canonical progeny of the helium-rich sub-population
  ($Y$$\approx$0.4, Norris \cite{norr04}; D'Antona et
  al. \cite{dabe05}) postulated to explain the observed split
  among the main-sequence stars of $\omega$\,Cen and NGC\,2808 (Bedin
  et al. \cite{bepi04}; Piotto et al. \cite{pivi05,pibe07}).  D'Antona
  \& Ventura (\cite{dave07}) propose a model in which stars born with
  some helium-enrichment experience non-canonical deep mixing on the
  red giant branch, which increases their helium abundance to values
  of $Y>0.5$, reaching values of 0.6\ldots0.7 for special extra-mixing
  formalisms (D'Antona, priv. comm). The concept of helium-enriched
  sub-populations in globular clusters has been extended to the point
  where some authors claim that most, if not all, globular clusters
  contain {\em highly} helium-enriched populations (e.g. D'Antona \&
  Caloi \cite{daca08}, Caloi \& D'Antona \cite{cada08}, [M\,3], Di
  Criscienzo et al. \cite{dida10} [NGC\,6397]).  On the other hand,
  Catelan et al. (\cite{cagr09}) use high-precision observations of
  M\,3 HB stars to rule out helium enhancement of $\Delta Y \gtrsim
  0.02$.  Villanova et al. (\cite{vipi09}) attempted to determine the
  helium abundance of a globular cluster with a very extended
  horizontal branch (NGC\,6752), but found no evidence of helium
  enhancement in HB stars between 8500\,K and 9000\,K\footnote{Fabbian
  et al. (\cite{fare05}) also presented helium abundances for two HB
  stars in NGC\,1904 that lie redward of the diffusion threshold, but
  their error bars of $\pm$0.3 dex prohibit any conclusion about
  helium enhancement.}. However, as their targets are Na-poor and
  O-rich and therefore unlikely to be helium enriched, this result is
  inconclusive. Sandquist et al. (\cite{sago10}) in a very extensive
  study of the globular cluster M\,13 found no evidence of the
  significant helium enrichment claimed by Caloi \& D'Antona
  (\cite{cada05}) and D'Antona \& Caloi (\cite{daca08}). 
  Portinari et al. (\cite{poca10}) suggested that the stellar
  evolution models of the lower main sequence do not predict the
  correct relation between effective temperature and helium content.
  Evolutionary models constructed for a canonical helium content
  suggest a helium-to-metal enrichment ratio of $\Delta Y/\Delta Z$=10
  for local stars on the lower main sequence. This ratio would imply a
  helium content significantly smaller than the primordial helium
  abundance ($Y$$\approx$0.1) for the most metal-poor local stars. On
  the other hand, empirical relations reproducing the observed
  behaviour of local lower main-sequence stars for $\Delta Y/\Delta
  Z$=2 suggest a lower helium enrichment for the globular clusters
  $\omega$\,Cen and NGC\,2808, thereby reducing the problems in
  achieving that enrichment.

For an excellent review of the r\^ole of helium enrichment in the
problem of the second parameter we refer to Gratton et
  al. (\cite{grca10}). They argue that a moderate helium enrichment of
  $\Delta$$Y < 0.1$ represents a third parameter (in addition to the
  second parameter, age). While $\omega$\,Cen is unfortunately not
  part of their study they analyse many of the blue-hook globular
  clusters mentioned earlier. This moderate value is also supported by
  the analysis of the red giant branch stars in 19 globular clusters
  from the Gratton et al. sample by Bragaglia et al. (\cite{brca10}).

Lee et al. (\cite{lejo05}) suggested that the blue hook stars
are the progeny of the proposed helium-rich main-sequence stars in
$\omega$\,Cen. D'Antona et al. (\cite{daca10}) proposed 
that two populations of very helium-rich HB stars ($Y$ = 0.8 and $Y$
= 0.65, corresponding to \loghe\ = 0 and $-$0.33, respectively), which
achieve their extreme abundances via extra mixing processes during the
red giant branch evolution of $Y$$\approx$0.4 stars, could
explain the observations of blue hook stars in $\omega$\,Cen.  If the blue
hook stars were to be explained {\em solely} by the helium-enrichment
scenario, their helium abundance should therefore not exceed
$Y$$\approx$0.8 and carbon should not be enriched at all.
Spectroscopic observations of the blue (and supposedly helium-rich)
main-sequence stars in $\omega$~Cen yielded a carbon abundance of
[C/M] = 0.0 (Piotto et al. \cite{pivi05}).  This carbon abundance will
decrease further as the stars ascend the red giant branch,
due to the extra-mixing process that occurs in metal-poor
red giants (Kraft \cite{kraf94}; Gratton et al. \cite{grsn00}).
Origlia et al. (\cite{orfe03}) confirmed that the RGB stars in
$\omega$\,Cen have the low $^{12}$C/$^{13}$C ratios ($\approx$4) and
low average carbon abundances ([C/Fe] = $-$0.2) expected from this
extra mixing.  Thus the helium-enrichment scenario predicts a carbon
abundance by mass in the blue hook stars of less than 0.1\%, i.e., at
least a factor of 10 smaller than the carbon abundance predicted by
the flash-mixing scenario. A UV study of five
massive globular clusters with blue hook stars by Brown et
al. (\cite{brsw10}) also showed that the faint luminosities observed for
these stars can only be explained by the late hot flasher
scenario. However, neither of the two scenarios can explain the range
of colours observed for blue hook stars, especially in the more
metal-rich globular clusters. Also Dalessandro et al. (\cite{dasa10})
are unable to reproduce the UV and optical photometry of the blue hook stars
in NGC\,2808 solely with helium enrichment.

Previous spectra of the blue hook stars in $\omega$\,Cen (Moehler et
al. \cite{mosw02}) and NGC\,2808 (Moehler et al.\ \cite{mosw04}) showed
that these stars are indeed both hotter and more helium-rich than the
canonical EHB stars. However, the blue hook stars show evidence for
considerable amounts of hydrogen. Unfortunately, due to the limited
resolution and signal-to-noise ratios (S/N) of the available
  data we have been unable to derive reliable
abundances for C and N. We could instead only state that the most
helium-rich stars appear to show some evidence of C/N
enrichment. We therefore started a project to obtain higher resolution
spectra of hot and extreme HB stars and blue hook stars in
$\omega$\,Cen. Our first results were published in Moehler et
al. (\cite{modr07}).

\section{Observations}\label{sec:obs}
\begin{figure}[h!]
\includegraphics [height=\columnwidth,angle=0]{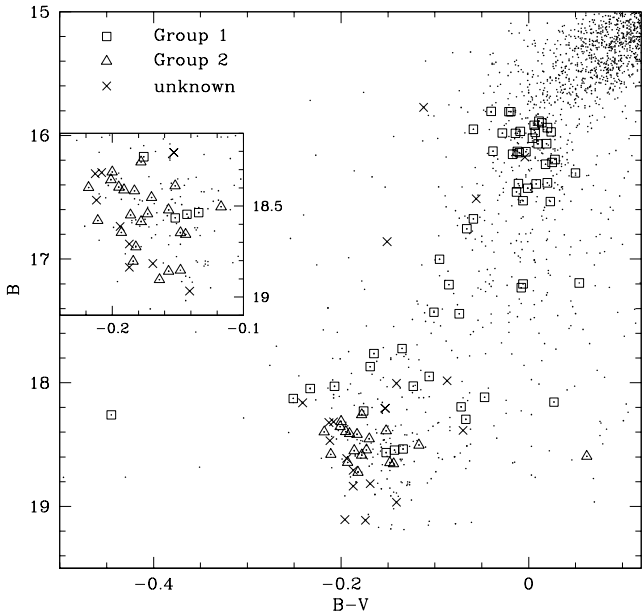}
\caption[]{$B, B-V$ colour-magnitude diagram with the FLAMES targets
  marked. 'Unknown' refers to targets where the spectra did not allow
  an analysis. See text for details.}\label{Fig:cmd} 
\end{figure}

We selected stars along the blue HB in $\omega$\,Cen from the
multi-band ($U,B,V,I$) photometry of Castellani et
  al. (\cite{cast07}), observed with the Wide Field Imager at the
2.2m\,MPG/ESO telescope.  These data together with multiband data
  from the Advanced Camera for Surveys onboard the Hubble Space
  Telescope provided the largest sample of HB stars ($\approx$3,200)
  ever collected for a globular cluster. Among them we
  concentrated on the stars at the faint end of the HB, which are the
  most likely ``blue hook'' candidates as shown by Moehler et
  al. (\cite{mosw02}, \cite{mosw04}). To avoid crowding
  problems, we tried to select only isolated stars. The astrometry was
  performed using the UCAC2 catalog (Zacharias et
  al. \cite{zacharias2004}), which does not cover the central crowded
  regions. However, thanks to the large field covered by the current
  data set the astrometric solution is based on
  $\approx$3\,000~objects with an rms error of 0\farcs06. The targets
  are marked in Fig.~\ref{Fig:cmd} and listed in
  Table~\ref{Tab:targets}.

\begin{table}
\caption[]{Target coordinates, heliocentric radial velocities, and 
    $B$ brightness}\label{Tab:targets}
\begin{tabular}{lllrr}
\hline
\hline
number & $\alpha_{2000}$ & $\Delta_{2000}$ & $B$ & RV \\
 & & & & [km s$^{-1}$] \\
\hline
 24609$^1$ & 13:25:34.3 & $-$47:29:50.0 & 18.708& 254.0\\
 25751$^1$ & 13:25:35.6 & $-$47:27:45.3 & 18.836& 197.8 \\
 26774 & 13:25:36.6 & $-$47:31:07.9     & 18.645& 213.0  \\
 29850 & 13:25:39.8 & $-$47:28:57.0     & 18.411& 240.0  \\
 30675 & 13:25:40.5 & $-$47:33:27.7     & 17.206& 220.0  \\
 31400 & 13:25:41.3 & $-$47:29:06.3     & 18.803& 227.6  \\
 32033$^2$ & 13:25:42.0 & $-$47:25:38.7 & 15.773& 258.6  \\ 
 35357 & 13:25:45.0 & $-$47:36:47.5     & 15.806& 233.0  \\
 35828 & 13:25:45.5 & $-$47:28:27.9     & 18.229& 199.5  \\
 36669 & 13:25:46.5 & $-$47:22:36.6     & 18.127& 205.6  \\
 36725 & 13:25:46.4 & $-$47:26:52.2     & 18.967& 241.0  \\
 39876$^1$ & 13:25:49.1 & $-$47:36:04.0 & 18.316& -- \\
 40846 & 13:25:50.0 & $-$47:29:45.3     & 18.417& 255.3  \\
 41074 & 13:25:50.1 & $-$47:32:06.3     & 18.521& 227.7  \\
 43148$^1$ & 13:25:51.9 & $-$47:31:39.4 & 18.468& 227.1 \\
 43520 & 13:25:52.3 & $-$47:22:57.8     & 18.580& 227.4  \\
 45556 & 13:25:53.8 & $-$47:34:48.7     & 18.398& 222.0  \\
 45715$^3$ & 13:25:54.0 & $-$47:29:07.8 & 16.512& --  \\
 45734$^1$ & 13:25:54.0 & $-$47:35:21.6 & 18.612& 255.4  \\
 51341 & 13:25:58.6 & $-$47:23:50.8     & 15.808& 207.1  \\
 51359 & 13:25:58.5 & $-$47:30:59.8     & 18.595&218.5  \\
 53022$^1$ & 13:25:59.8 & $-$47:37:45.8 & 18.322&266.7  \\
 53367$^4$ & 13:26:00.1 & $-$47:35:17.5 & 19.112&209.1  \\
 53945 & 13:26:00.7 & $-$47:29:28.3     & 18.724& 203.7  \\
 54733 & 13:26:01.4 & $-$47:22:36.8     & 18.858& 217.4  \\
 55158 & 13:26:01.5 & $-$47:35:43.8     & 18.396& 251.7  \\
 56896 & 13:26:03.0 & $-$47:20:29.3     & 18.258& 236.2  \\
 58774 & 13:26:04.3 & $-$47:31:32.7     & 18.296& 216.9  \\
 59125 & 13:26:04.6 & $-$47:30:21.9     & 15.807& 231.0  \\
 59786 & 13:26:05.2 & $-$47:21:44.9     & 18.585& 229.2  \\
 60820 & 13:26:05.8 & $-$47:29:25.2     & 18.029& 226.7  \\
 65373 & 13:26:08.8 & $-$47:37:12.6     & 18.355& 232.3  \\
 66104$^3$ & 13:26:09.4 & $-$47:29:57.5 & 15.593& --  \\
 66703$^1$ & 13:26:09.8 & $-$47:25:10.7 & 18.162& 187.7 \\
 67933$^1$ & 13:26:10.5 & $-$47:39:09.9 & 18.896& -- \\
 69373 & 13:26:11.5 & $-$47:27:51.5     & 17.870& 219.3  \\
 71099 & 13:26:12.6 & $-$47:36:13.8     & 19.107& 264.3	\\
 72787 & 13:26:13.8 & $-$47:30:58.2     & 15.917& 211.5  \\
 74069$^4$ & 13:26:14.6 & $-$47:26:47.5 & 16.859& 229.4  \\
 75364 & 13:26:15.5 & $-$47:25:03.0     & 18.904& 220.2  \\ 
 75981$^4$ & 13:26:15.9 & $-$47:28:09.2 & 18.313& 211.5  \\
 75993$^4$ & 13:26:15.9 & $-$47:28:30.4 & 18.206& 220.3  \\
 80690 & 13:26:19.0 & $-$47:20:20.3     & 18.453& 217.5  \\
 80711 & 13:26:18.9 & $-$47:25:09.1     & 15.898& 220.4  \\
 81395 & 13:26:19.4 & $-$47:22:17.1     & 15.884& 220.1  \\
 81722 & 13:26:19.6 & $-$47:28:49.0     & 16.304& 239.1  \\
 82860 & 13:26:20.4 & $-$47:20:14.1     & 15.983& 228.7  \\
 86429 & 13:26:22.6 & $-$47:30:52.9     & 17.201& 206.1  \\
 87161 & 13:26:23.0 & $-$47:39:36.6     & 18.543& 236.2  \\
 87175 & 13:26:23.1 & $-$47:26:10.3     & 16.132& 235.5  \\
 87734 & 13:26:23.5 & $-$47:21:54.5     & 16.140& 215.5  \\
 89495 & 13:26:24.8 & $-$47:23:25.2     & 15.937& 207.2  \\
 92333$^4$ & 13:26:26.7 & $-$47:31:05.7 & 17.193& 213.9  \\
 93516 & 13:26:27.6 & $-$47:21:13.7     & 16.022& 226.0  \\
 95401 & 13:26:28.9 & $-$47:36:20.8     & 18.536& 212.2  \\
 96242 & 13:26:29.6 & $-$47:20:44.5 & 16.176& 216.6  \\
\hline
 \multicolumn{2}{l}{$^1$: very noisy spectra} &
 \multicolumn{3}{l}{$^2$: only 2 spectra extracted}\\
 \multicolumn{2}{l}{$^3$: no spectra extracted} &
 \multicolumn{3}{l}{$^4$: G-band and/or \ion{Fe}{i} 4325\AA\ visible}\\
\end{tabular}
\end{table}
\setcounter{table}{0}
\begin{table}[!h]
\caption[]{{\bf (cont'd)} Target coordinates, heliocentric radial
  velocities, and $B$ brightness}
\begin{tabular}{llrrr}
\hline
\hline
number & $\alpha_{2000}$ & $\Delta_{2000}$ & $B$ & RV \\
 & & & & [km s$^{-1}$] \\
\hline
102600 & 13:26:34.1 & $-$47:20:57.5     & 17.723& 213.0  \\
102850 & 13:26:34.3 & $-$47:23:20.3     & 16.383& 233.5  \\
103563 & 13:26:34.7 & $-$47:40:24.8     & 18.028& 205.2  \\
111785 & 13:26:40.5 & $-$47:20:44.5     & 19.141& 239.2  \\
112475 & 13:26:40.9 & $-$47:18:42.6     & 18.504& 232.6  \\
114491 & 13:26:42.3 & $-$47:37:29.5     & 16.528& 223.3  \\
115087 & 13:26:42.7 & $-$47:36:33.8     & 16.534& 235.3  \\
115194 & 13:26:42.8 & $-$47:21:18.0     & 18.117& 229.0  \\
120119 & 13:26:46.2 & $-$47:22:49.7     & 18.852& 230.2  \\
120901$^4$ & 13:26:46.8 & $-$47:33:29.0 & 18.987& 216.5  \\
124014$^4$ & 13:26:48.9 & $-$47:33:30.1 & 19.036& 226.0 \\
125302 & 13:26:49.7 & $-$47:21:52.3     & 16.233& 228.6  \\
126350 & 13:26:50.5 & $-$47:22:05.8     & 17.231& 227.3  \\
126892 & 13:26:50.9 & $-$47:37:10.5     & 18.389& 229.4  \\
130310 & 13:26:53.1 & $-$47:36:13.0     & 15.981& 244.9  \\
130831 & 13:26:53.5 & $-$47:33:09.8     & 16.195& 244.4  \\
133846 & 13:26:55.3 & $-$47:35:57.4     & 16.129& 221.5  \\
135227$^4$ & 13:26:56.2 & $-$47:34:36.8 & 18.816& 205.5 \\
137299 & 13:26:57.4 & $-$47:22:59.5     & 17.949& 217.3  \\
141008 & 13:26:59.8 & $-$47:21:14.0     & 18.646& 226.9  \\
144749$^4$ & 13:27:02.2 & $-$47:25:45.2 & 18.386& 238.2  \\
145078$^3$ & 13:27:02.4 & $-$47:25:10.6 & 17.984& --  \\
147746 & 13:27:04.2 & $-$47:35:02.3     & 15.967& 226.3  \\
147880 & 13:27:04.3 & $-$47:34:25.4     & 15.950& 234.0  \\
148641$^4$ & 13:27:04.7 & $-$47:23:02.8 & 19.815& 210.8  \\
156459 & 13:27:10.2 & $-$47:27:33.8     & 15.975& 210.9  \\
161310$^1$ & 13:27:13.7 & $-$47:39:23.6 & 18.047& 234.6  \\
162839 & 13:27:14.6 & $-$47:24:07.4     & 16.390& 224.6  \\
164808 & 13:27:16.1 & $-$47:34:01.6     & 18.549& 222.2  \\
165244 & 13:27:16.5 & $-$47:39:28.4     & 15.972& 232.2  \\
169814 & 13:27:19.8 & $-$47:26:55.3     & 16.392& 233.5  \\
170215 & 13:27:20.2 & $-$47:36:06.5     & 16.068& 221.7  \\
170450$^1$ & 13:27:20.5 & $-$47:38:41.8 & 18.005& 236.4  \\
171696 & 13:27:21.3 & $-$47:27:18.0     & 16.067& 243.3  \\
172332 & 13:27:21.7 & $-$47:23:44.6     & 16.427& 242.6  \\
174389 & 13:27:23.6 & $-$47:32:31.0     & 16.152& 236.0  \\
177314 & 13:27:26.0 & $-$47:32:42.6     & 17.442& 219.8  \\
180375 & 13:27:28.7 & $-$47:30:34.5     & 16.218& 214.2  \\
180714$^4$ & 13:27:29.1 & $-$47:34:59.6 & 18.156& 221.3  \\
181678 & 13:27:29.8 & $-$47:24:21.8     & 18.599& 244.2  \\
182005 & 13:27:30.3 & $-$47:32:23.8     & 16.457& 228.3  \\
182772 & 13:27:31.0 & $-$47:27:16.9     & 17.764& 237.3  \\
183592 & 13:27:32.0 & $-$47:28:55.2     & 18.195& 233.3  \\
186476 & 13:27:36.2 & $-$47:32:41.8     & 18.545& 236.3  \\
187534 & 13:27:37.9 & $-$47:30:30.9     & 18.260& 217.9  \\
188882 & 13:27:40.4 & $-$47:34:02.4     & 17.001& 219.0  \\
189080 & 13:27:40.8 & $-$47:31:53.5     & 18.511& 235.6  \\
190398 & 13:27:43.3 & $-$47:27:00.5     & 18.656& 233.8  \\
190635 & 13:27:43.8 & $-$47:27:23.0     & 16.675& 239.3  \\
191111 & 13:27:44.7 & $-$47:25:19.8     & 18.565& 225.0  \\
191969 & 13:27:46.7 & $-$47:33:15.5     & 16.127& 227.5  \\
193486 & 13:27:50.3 & $-$47:32:17.6     & 16.755& 224.4  \\
194383 & 13:27:52.6 & $-$47:27:39.0     & 17.430& 217.3  \\
\hline
\end{tabular}
\end{table}

The spectroscopic data were obtained in 2005 (4 observations) and in
2006 (5 observations) in Service Mode using the MEDUSA mode of the
multi-object fiber spectrograph FLAMES$+$GIRAFFE on the UT2 Telescope
of the VLT. We used the low spectroscopic resolution mode with the
spectral range 3964\AA\ -- 4567\AA\ (LR2, R = 6400) and observed
spectra for a total of 109 blue hook and canonical blue HB/EHB star
candidates (see Table~\ref{Tab:targets}) and for 17 sky background
positions. Each observation had an exposure time of 2550 seconds to
keep the total execution time of the observing block shorter than one
  hour. Table~\ref{Tab:obs} lists the observing
  conditions. Unfortunately, only 9 of 20 planned observations could
  be obtained, which limits the S/N especially for the fainter stars.
\begin{table}
\caption[]{Observing log. The seeing given is that measured by the
  DIMM. The true seeing at the UT is often better.}\label{Tab:obs}
\begin{tabular}{llllr}
\hline
\hline
start of & airmass & seeing & \multicolumn{2}{c}{moon}\\
exposure & & & illum. & dist. \\
 UT & & [\arcsec] & & [$^\circ$]\\ 
\hline
2005-04-04T02:53:11.238 & 1.204 & 0.61 & 0.264 & 89.9 \\
2005-04-10T07:55:31.608 & 1.444 & 0.49 & 0.027 & 146.3 \\
2005-04-11T08:03:46.508$^1$ & 1.420 & 1.09 & 0.071 & 145.4 \\
2005-04-16T02:15:02.180 & 1.187 & 0.88 & 0.445 & 109.5 \\
2005-04-17T07:30:33.470 & 1.455 & 0.77 & 0.560 & 97.4 \\
2006-04-02T05:00:55.927 & 1.089 & 0.67 & 0.181 & 142.9\\
2006-06-15T23:13:58.802 & 1.112 & 0.62 & 0.780 & 97.2\\
2006-06-28T23:11:12.410 & 1.089 & 0.59 & 0.104 & 88.1\\
2006-06-30T01:37:27.558 & 1.245 & 1.08 & 0.177 & 76.8\\
2006-07-04T00:55:50.945 & 1.190 & 0.78 & 0.528 & 40.8\\
2006-08-09T23:31:50.834 & 1.349 & 1.59 & 0.994 & 101.2\\
2006-08-09T23:25:30.528 & 1.261 & 1.51 & 0.994 & 101.1\\
\hline
\multicolumn{5}{l}{$^1$: exposure aborted after 11 minutes}\\
\hline
\end{tabular}
\end{table}

\section{Data reduction}\label{sec:reduction}
For our first analysis presented in Moehler et al. (\cite{modr07}), we
used the ESO pipeline-reduced data. Unfortunately the pipeline version
used for those data did not correct for the bright spot seen at the
upper right corner of the GIRAFFE CCD. Therefore we reduced the data
again using the Geneva pipeline
GirBLDRS\footnote{http://girbldrs.sourceforge.net/} (version 1.11).
To remove the bright spot from the data we obtained raw dark frames
from the ESO
archive\footnote{http://archive.eso.org/wdb/wdb/eso/giraffe/form} for
the time range covered by our data. First we created master bias
frames by averaging the five bias frames that had been obtained for
each observation. Then the three dark frames observed on a given date
were bias corrected and averaged with cosmic ray rejection. To reduce
the noise in the dark frames, we smoothed them with a 2$\times$2 pixel
box filter. We then used the three flat fields observed for each night
to determine the positions of the spectra. We did not correct these
data for dark current, as the bright spot showed no negative effect on
the detection of the spectra. Afterwards we derived the full
wavelength solution from the ThAr arc frames observed for each
date. Using this solution, we finally extracted and rebinned the
science data. At this step we included the smoothed master dark
frames. The wavelength calibration was adjusted using the
simultaneously observed ThAr spectra. As the optimal extraction
produced spectra of lower signal-to-noise than conventional averaging
we used the average option to integrate the flux for each spectrum. We
also divided the spectra by extracted flat-field spectra to perform a
first correction for the CCD sensitivity variations. Unfortunately, no
flux standard stars are observed by GIRAFFE in the MEDUSA mode, which
would permit at least a relative flux calibration to be obtained.

For each exposure, we subtracted the median of the spectra from the sky
fibers from the extracted spectra of the target stars. We corrected
all spectra for barycentric motion. During this correction, we noted
that we had made a sign error when performing this correction for the
data published by Moehler et al. (\cite{modr07}). This error led to a
smearing of the line profiles in the averaged spectra.

\begin{figure}[h!]
\includegraphics [height=1.1\columnwidth,angle=270]{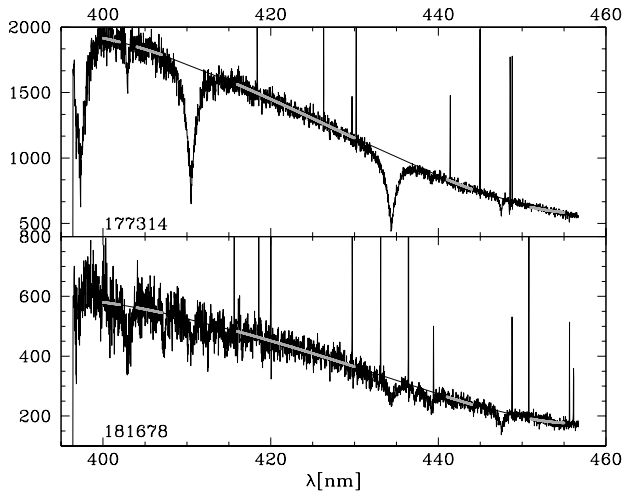}
\caption[]{Spectra and fits to them from April 2, 2006, for a
  helium-poor star (top) and a helium-rich star (bottom). The regions
  used for the fit are marked in grey.}\label{Fig:norm}
\end{figure}

When comparing spectra observed at different dates, we found that the
slope of the extracted spectra tends to vary from one observation to
the next. To average the spectra, we therefore first normalized
the individual spectra. To achieve this we fitted a 5$^{th}$
order polynomial to regions free from strong lines (4000--4020\,\AA,
4040--4070\,\AA, 4160--4300\,\AA, 4410--4440\,\AA, and
4510--4550\,\AA). In Fig.~\ref{Fig:norm}, we show example fits for
helium-poor and helium-rich stars.

\section{Radial velocities}\label{sec:RV}
\begin{figure}[h!]
\includegraphics [height=\columnwidth,angle=0]{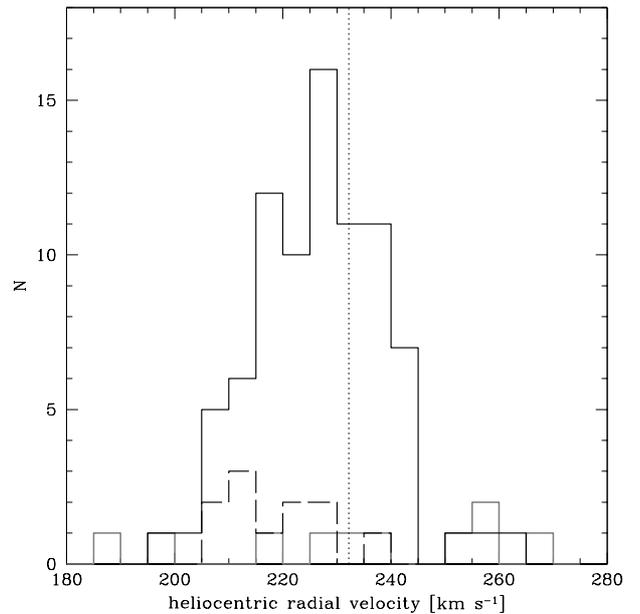}
\caption[]{The distribution of heliocentric radial velocities. The
  solid black line shows the distribution for spectra with no notation
  in Table~\ref{Tab:targets}, the dashed black line shows the
  distribution for spectra showing a G-band and the solid grey line
  shows the distribution for noisy spectra. The
  heliocentric radial velocity of $\omega$\,Cen is marked by the dotted
  line at 232.2\,km s$^{-1}$ (Harris \cite{harr96}).}\label{Fig:histo}
\end{figure}

After the barycentric correction, the observed spectra were -- in a
first step -- cross-correlated with synthetic spectra roughly matching
the stellar parameters, i.e. helium abundance, surface gravity, and
effective temperature. Only regions of hydrogen or helium lines were
selected prior to the cross-correlation. The peak of the
cross-correlation function was then fitted with a Gaussian
function to determine the radial velocity to sub-resolution accuracy. The
velocity-corrected spectra were then co-added and fitted with
synthetic model atmospheres (see Sect.~\ref{sec:analysis}). In a
second step, the best-fit synthetic spectra were then used to
repeat the cross-correlation. Figure~\ref{Fig:histo} shows the
distribution of the radial velocities. The median radial velocity of
the 83 ``clean'' target stars (i.e. with sufficient signal and without
G-band\footnote{The presence of a G-band in a hot star's spectrum
indicates an optical or physical binary with a cool companion.}) of
226.8\,km\,s$^{-1}$ derived this way differs by 5.4\,km\,s$^{-1}$ from
the accepted value for $\omega$\,Cen of 232.2\,km\,s$^{-1}$ (Harris
\cite{harr96}). Using a more sophisticated approach that 
accounts for the uncertainties in the individual measurements and fits
a Gaussian to the velocity distribution of {\em all} target stars with
measurable radial velocities, we derive a marginally lower heliocentric
radial velocity of 226.5\,km\,s$^{-1}$ and a standard deviation $\sigma$ of
17.4\,km\,s$^{-1}$.

\begin{figure}[h!]
\includegraphics [height=\columnwidth,angle=0]{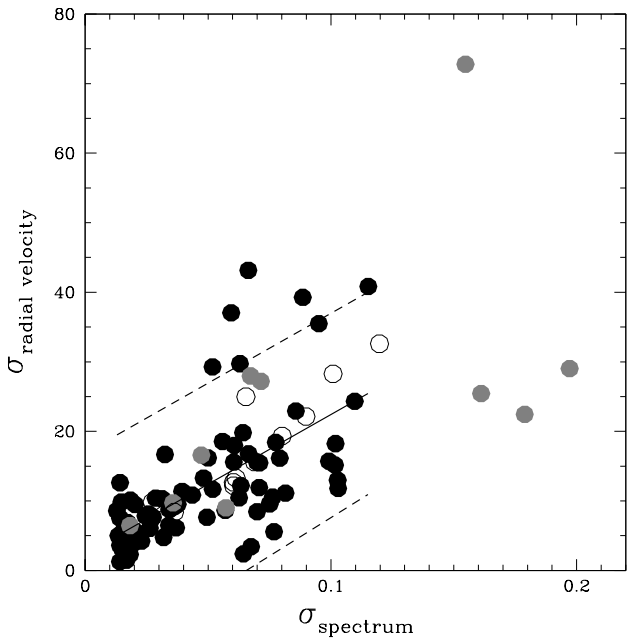}
\caption[]{The scatter in the radial velocities from individual spectra
  versus the signal-to-noise ratio of the spectra. The solid black
  dots are from spectra with no annotation in Table~\ref{Tab:targets}
  (i.e. good spectra), the grey dots are from noisy spectra, and the
  open symbols refer to spectra containing a G band.}\label{Fig:SigmaRV}
\end{figure}

The mean and standard deviation of the radial velocity were determined
from the up to nine individual spectra for each object. This standard
deviation was compared to the standard deviation in the spectral
flux, calculated from a continuum region between 4150 and
4250\,\AA. By comparing these two quantities we found a
linear dependence of the scatter in the radial velocity measurements
for each star on the quality of the spectra (see
Fig.~\ref{Fig:SigmaRV}).  The plot includes the linear fit to the good
data as well as the 2\,$\sigma$-limits.  Stars with a significantly
higher scatter might be primary stars within a binary, where the
scatter represents orbital motions with periods smaller than the
duration of the observing campaign. Ignoring the few stars with
very poor spectra, i.e. with a spectral standard deviation in
  flux above 0.1, we identify five objects that may be
binary members.

For each target star, we fitted sine curves by stepping through
test periods from 1 to 230 days adopting the standard deviation
of the radial velocity measurements as the amplitude, the mean
as the velocity offset, and an arbitrary phase as a
starting approximation. The amplitude, period, and phase were allowed
to vary during the fit procedure. The period belonging to the fit with
the lowest standard deviation in the residuals was taken as
possible binary period. For each object, we then created a series of
radial velocity curves with identical times but randomly distributed
radial velocity measurements with a standard deviation equaling
that of the RV measurements. Each of these curves was then
fitted with the previously determined possible period. We counted the
fits with amplitudes smaller than that of the original light curve to
derive a false alarm probability. Out of our sample, 21 
objects have a false alarm probability below 5\%. However, none of
these experience a significant reduction in standard
deviation between the original RV measurements and the residuals 
for the best-fit period. Among these 21, there is also no overlap
with the five objects showing an excess RV standard deviation with
respect to the spectral S/N. We therefore conclude that we cannot
reliably identify spectroscopic binaries within our sample.

After verifying that there were no significant radial velocity
variations, we then averaged the individual spectra for each star,
excluding at each wavelength the two highest and the two lowest data
points. Usually this meant that 5 data points were used to
  compute the
average. However, some stars could not be extracted in all exposures
or had data of too-low S/N to determine a radial velocity. In these cases,
the number of data points available for averaging was obviously
smaller. This affected the stars 36725 (2 spectra), 45734 (3 spectra),
190635 (4 spectra), 43148, 53022, 53945, 66703, 71099 (5 spectra
each), 55158, 75364 164808 (6 spectra each), 24609, 170450, 182005 (7
spectra each), and 25751, 80690, 112475, 162839, 180375, 180714,
182772, 188882, 190398, and 191111 (8 spectra each). To
check if the stars with fewer spectra had significantly lower S/N we
fitted the S/N of the spectra determined between 420\,nm and
423\,nm versus the $B$
magnitude for stars with nine spectra available for averaging. 
We found that the S/N of the stars with fewer spectra is within
$\pm$2$\sigma$ ($\pm$50\%) of this fit for all stars with
$B$$\lesssim$18.4. For fainter stars (24609, 25751, 43148, and 45734),
the lack of spectra yields averaged spectra with a S/N below 12 in
the studied region, which is the empirical limit for a meaningful
analysis within this data set. There are, however, also two objects
with nine spectra that are below this limit: 39876 and 67933. 

\section{Analysis}\label{sec:analysis}
\subsection{Atmospheric parameters}\label{Sec:parameters}

To derive effective temperatures, surface gravities, and
helium abundances, we fitted the Balmer lines H$_\gamma$ and H$_\delta$
(and/or the \ion{He}{ii} lines at these positions) and the \ion{He}{i}
lines 4026~\AA, 4388~\AA, and 4471~\AA. 

To establish the best fit to the observed spectra, we used the
routines developed by Bergeron et al. (\cite{besa92}) and Saffer et
al. (\cite{sabe94}), as modified by Napiwotzki et al. (\cite{nagr99}),
which employ a $\chi^2$ test. The $\sigma$ necessary for the
calculation of $\chi^2$ is estimated from the noise in the continuum
regions of the spectra. The fit program normalizes both the model {\em
and} observed spectra using the same points for the continuum
definition.

Recent tests have shown, however, that these fit routines
underestimate the {\em formal} errors by at least a factor of 2
(Napiwotzki priv. comm.). We therefore provide formal errors
multiplied by 2 to account for this effect. In addition, the errors
provided by the fit routine do not include possible systematic errors
caused by, e.g., flat-field inaccuracies or imperfect sky
subtraction. The true errors in \teff\ are probably close to 5\% at
least, and the true errors in \logg\ are probably about 0.1.

The spectra were fitted with various model atmospheres.  As the late
hot flasher scenario predicts enrichment in carbon and nitrogen, an
extensive grid of {\em non-LTE line-blanketed} model atmospheres has
been produced with the NLTE model atmosphere code
TLUSTY\footnote{http://nova.astro.umd.edu} (Hubeny \& Lanz
\cite{hula95}). These model atmospheres allow for departures from LTE for
1132 explicit levels and superlevels of 52 ions (H, He, C, N, O, Ne,
Mg, Al, Si, P, S, Fe). A detailed description of the model atoms and
the source of the atomic data can be found in Lanz~\& Hubeny
(\cite{lahu03}, \cite{lahu07}). The model grid covers the range of
stellar parameters typical of EHB stars: 20\,000\,K $\le$\teff$\le$
50\,000\,K (step of 2500\,K), 4.75$\le$\logg$\le$6.5 (step of 0.25
dex), and $-3\le$\loghe$\le+$2 (step of 1 dex) at a microturbulent
velocity, $\xi$ = 5 km s$^{-1}$.  For each model atmosphere with
\loghe $\ge-1$, we calculated a second model with {\em enriched
content of carbon and nitrogen} (marked by $^{\rm C}$ in
Table~\ref{Tab:Par_Rich}) following the prediction of the ``flash
mixing'' scenario, adopting mass fractions of 3\% and 1\% for carbon
and nitrogen, respectively (Lanz et~al. \cite{labr04}). We adopted
scaled-solar abundances at $\omega$\,Cen's dominant
metallicity ([Fe/H] = $-$1.5). This abundance ratio by numbers was
kept the same for all models, including helium-rich models, which
implies that the heavy element mass fraction differs for models with
different helium (and C, N) content. We emphasize, however, that the
abundance of iron-peak elements in EHB stellar photospheres is unknown
and probably affected by diffusion processes. Furthermore, the low
abundance of heavy elements limits the effect of metal line blanketing
on the atmospheric structure and the predicted emergent spectrum.
Therefore, the resulting uncertainty in our analysis caused
  by assuming the
same [Fe/H] value remains small. Once the atmospheric
structure of each model atmosphere converged, we calculated detailed
emergent spectra in the $\lambda\lambda$3800-4600\,\AA\ range with the
spectrum synthesis code, SYNSPEC, using the NLTE populations
calculated by TLUSTY.

For the helium-poor stars above 20\,000\,K, we also used the TLUSTY
models. For the cooler stars, we used {\em metal-rich
helium-poor LTE} models (Moehler et al. \cite{mosw00}).

Using the atmospheric parameters, a distance modulus of $(m-M)_0 =
\magpt{13}{45}$, and an interstellar absorption of $A_V =
\magpt{0}{47}$, we derived masses for our target stars as described in
Moehler et al. (\cite{mosw00}). The masses for the helium-rich stars
are somewhat underestimated as we used theoretical brightness values
for solar-helium atmospheres, which are brighter in the optical range
than helium-rich atmospheres.

\begin{figure}[h!]
\includegraphics [height=1.0\columnwidth,angle=270]{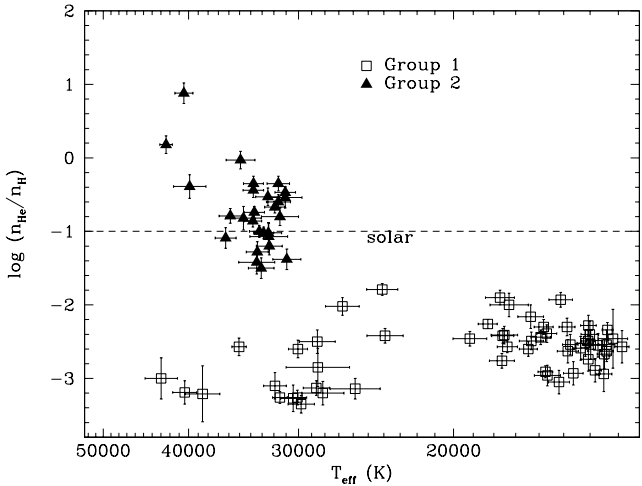}
\caption[]{The effective temperatures and helium abundances by number
 derived for our target stars (formal errors multiplied by 2, see text
 for details).  Helium-poor and helium-rich stars are
 marked by open squares and filled triangles, respectively. The stars with
 super-solar helium are shown with the parameters derived from
 models without C/N enhancement. The dashed line indicates the solar
 helium abundance}\label{Fig:Teff_loghe}
\end{figure}

\begin{table}
\caption[]{Atmospheric parameters for stars with sub-solar helium abundances}\label{Tab:Par_Poor}
\begin{tabular}{lrrrr}
\hline
\hline
number & $\chi^2$ & \teff & \logg & \loghe \\
\hline
29850 & 1.105 & 33400$\pm$1000 & 5.87$\pm$0.18 & $-$1.28$\pm$0.14\\
30675 &1.583 & 16400$\pm$\ \ 600 & 4.54$\pm$0.10 & $-$2.16$\pm$0.16 \\
32033 &1.227 & 17000$\pm$\ \ 600 & 4.09$\pm$0.10 & $-$1.52$\pm$0.10 \\
35357 &1.137 & 13200$\pm$\ \ 500 & 3.99$\pm$0.12 & $-$2.46$\pm$0.40 \\
35828 & 1.030 & 29800$\pm$1000 & 5.54$\pm$0.18 & $-$3.35$\pm$0.12\\
36669 & 1.300 & 38600$\pm$1800 & 5.68$\pm$0.20 & $-$3.21$\pm$0.38\\
40846 & 1.070 & 32400$\pm$1600 & 5.52$\pm$0.26 & $-$1.07$\pm$0.18\\
45556 & 1.271 & 33100$\pm$1100 & 5.68$\pm$0.20 & $-$1.50$\pm$0.14\\
51341 &1.796 & 12900$\pm$\ \ 200 & 3.90$\pm$0.06 & $-$2.57$\pm$0.22 \\
51359 & 1.310 & 32400$\pm$1100 & 5.62$\pm$0.18 & $-$1.20$\pm$0.12\\
58774 & 1.354 & 25900$\pm$1700 & 5.27$\pm$0.20 & $-$3.14$\pm$0.14\\
59125 &2.350 & 13400$\pm$\ \ 300 & 3.96$\pm$0.06 & $-$2.54$\pm$0.20\\
60820 & 1.091 & 40400$\pm$1300 & 5.46$\pm$0.12 & $-$3.19$\pm$0.16\\
65373 &1.062 & 33200$\pm$\ \ 900 & 5.69$\pm$0.14 & $-$1.00$\pm$0.08 \\
69373 & 1.122 & 30100$\pm$7800 & 5.48$\pm$0.12 & $-$3.26$\pm$0.10\\
72787 &1.182 & 13700$\pm$\ \ 300 & 4.05$\pm$0.06 & $-$2.55$\pm$0.16 \\
75981 & 1.056 & 33500$\pm$1600 & 5.42$\pm$0.26 & $-$1.42$\pm$0.16\\
80711 &2.113 & 13400$\pm$\ \ 200 & 4.03$\pm$0.06 & $-$2.63$\pm$0.14 \\
81395 &1.227 & 13400$\pm$\ \ 200 & 4.06$\pm$0.04 & $-$2.34$\pm$0.10 \\
81722 &2.376 & 17700$\pm$\ \ 700 & 4.18$\pm$0.10 & $-$1.90$\pm$0.10 \\
82860 &2.789 & 14900$\pm$\ \ 300 & 4.11$\pm$0.06 & $-$2.30$\pm$0.12 \\
86429 & 1.585 & 28200$\pm$1600 & 4.86$\pm$0.18 & $-$3.20$\pm$0.16\\
87175 &2.200 & 15700$\pm$\ \ 300 & 4.16$\pm$0.06 & $-$2.91$\pm$0.08 \\
87734 &2.588 & 14100$\pm$\ \ 300 & 4.13$\pm$0.06 & $-$2.74$\pm$0.16 \\
89495 &1.877 & 14000$\pm$\ \ 300 & 4.00$\pm$0.06 & $-$2.54$\pm$0.20\\
93516 &2.920 & 14300$\pm$\ \ 300 & 4.06$\pm$0.06 & $-$2.59$\pm$0.16 \\
95401 & 1.257 & 31900$\pm$1000 & 5.68$\pm$0.20 & $-$3.10$\pm$0.18\\
96242 &1.826 & 15900$\pm$\ \ 400 & 4.15$\pm$0.06 & $-$2.44$\pm$0.10\\
102600 & 1.172 & 28600$\pm$1000 & 5.25$\pm$0.14 & $-$3.13$\pm$0.10\\
102850 &1.852 & 16300$\pm$\ \ 500 & 4.23$\pm$0.08 & $-$2.49$\pm$0.12 \\
103563 & 1.249 & 26700$\pm$1200 & 5.30$\pm$0.14 & $-$2.02$\pm$0.12\\
112475 & 1.272 & 30900$\pm$1100 & 5.76$\pm$0.20 & $-$1.38$\pm$0.14\\
114491 &1.848 & 17400$\pm$\ \ 500 & 4.44$\pm$0.06 & $-$2.57$\pm$0.08 \\
115087 &1.826 & 18300$\pm$\ \ 500 & 4.51$\pm$0.06 & $-$2.26$\pm$0.06 \\
115194 & 1.351 & 30000$\pm$8000 & 5.56$\pm$0.12 & $-$2.60$\pm$0.12\\
125302 &2.068 & 14700$\pm$\ \ 300 & 4.10$\pm$0.06 & $-$2.54$\pm$0.14 \\
126350 & 1.325 & 24100$\pm$1000 & 4.99$\pm$0.12 & $-$1.79$\pm$0.08\\
130310 &3.165 & 14100$\pm$\ \ 200 & 4.05$\pm$0.06 & $-$2.49$\pm$0.14 \\
130831 &1.298 & 15700$\pm$\ \ 500 & 4.09$\pm$0.10 & $-$2.96$\pm$0.14 \\
133846 &2.828 & 14200$\pm$\ \ 300 & 4.05$\pm$0.06 & $-$2.52$\pm$0.18 \\
137299 & 1.144 & 31500$\pm$6500 & 5.48$\pm$0.12 & $-$3.26$\pm$0.08\\
141008 & 1.010 & 36300$\pm$1000 & 5.92$\pm$0.16 & $-$1.09$\pm$0.14\\
147746 &2.120 & 13500$\pm$\ \ 300 & 3.97$\pm$0.06 & $-$2.67$\pm$0.20 \\
147880 &2.743 & 14000$\pm$\ \ 300 & 3.96$\pm$0.06 & $-$2.40$\pm$0.18 \\
156459 &2.095 & 15100$\pm$\ \ 500 & 4.06$\pm$0.10 & $-$1.93$\pm$0.10 \\
162839 &1.616 & 15700$\pm$\ \ 400 & 4.15$\pm$0.06 & $-$2.39$\pm$0.12 \\
164808 & 1.489 & 32500$\pm$1100 & 5.93$\pm$0.18 & $-$1.02$\pm$0.14\\
165244 &3.106 & 13500$\pm$\ \ 300 & 4.04$\pm$0.06 & $-$2.94$\pm$0.24 \\
169814 &2.328 & 15200$\pm$\ \ 400 & 4.26$\pm$0.08 & $-$3.05$\pm$0.16 \\
170215 &2.409 & 13800$\pm$\ \ 200 & 4.10$\pm$0.06 & $-$2.89$\pm$0.16 \\
171696 &2.158 & 14600$\pm$\ \ 400 & 4.05$\pm$0.08 & $-$2.93$\pm$0.16 \\
172332 &2.145 & 16500$\pm$\ \ 400 & 4.36$\pm$0.06 & $-$2.60$\pm$0.10 \\
174389 &1.603 & 14800$\pm$\ \ 400 & 4.10$\pm$0.08 & $-$2.63$\pm$0.16 \\
177314 &1.109 & 19200$\pm$\ \ 900 & 4.89$\pm$0.10 & $-$2.46$\pm$0.10 \\
180375 &2.177 & 17600$\pm$\ \ 500 & 4.21$\pm$0.06 & $-$2.42$\pm$0.10 \\
182005 &1.628 & 15800$\pm$\ \ 400 & 4.25$\pm$0.06 & $-$2.30$\pm$0.10 \\
182772 & 1.093 & 35100$\pm$7100 & 5.50$\pm$0.12 & $-$2.57$\pm$0.12\\
183592 & 1.387 & 28600$\pm$1400 & 5.48$\pm$0.16 & $-$2.50$\pm$0.16\\
186476 & 1.048 & 30400$\pm$1000 & 5.63$\pm$0.18 & $-$3.27$\pm$0.18\\
\hline
\end{tabular}
\end{table}
\setcounter{table}{2}
\begin{table}
\caption[]{{\bf cont'd} Atmospheric parameters for stars with sub-solar helium abundances}
\begin{tabular}{lrrrr}
\hline
\hline
number & $\chi^2$ & \teff & \logg & \loghe \\
\hline
187534 & 1.090 & 43000$\pm$1800 & 5.89$\pm$0.20 & $-$3.00$\pm$0.28\\
188882 &1.375 & 17600$\pm$\ \ 600 & 4.62$\pm$0.08 & $-$2.76$\pm$0.10 \\
190635 &1.988 & 17300$\pm$900 & 4.45$\pm$0.14 & $-$2.00$\pm$0.12 \\
191111 & 0.964 & 28500$\pm$2400 & 5.50$\pm$0.32 & $-$2.85$\pm$0.26\\
191969 &1.884 & 14100$\pm$\ \ 300 & 4.12$\pm$0.06 & $-$2.28$\pm$0.14 \\
193486 &1.117 & 17500$\pm$\ \ 600 & 4.61$\pm$0.08 & $-$2.41$\pm$0.12 \\
194383 & 1.083 & 23900$\pm$1200 & 5.06$\pm$0.14 & $-$2.42$\pm$0.10\\
\hline
\end{tabular}
\end{table}

\begin{table}[!h]
\caption[]{Atmospheric parameters for stars with super-solar helium
  abundances as derived with TLUSTY atmospheres. $^{\rm C}$ indicates
the use of C/N enhanced model atmospheres, whereas no notation
indicates the use of model atmospheres using the cluster metallicity.}
\label{Tab:Par_Rich}
\begin{tabular}{lrrrr}
\hline
\hline
number & $\chi^2$ & \teff & \logg & \loghe \\
\hline
26774 &1.128 & 31100$\pm$1400 & 5.97$\pm$0.26 & $-$0.54$\pm$0.14 \\
26774$^{\rm C}$ &1.083 & 30300$\pm$1500 & 6.03$\pm$0.28 & $-$0.72$\pm$0.14 \\[1mm]
31400 &1.338 & 31600$\pm$\ \ 900 & 5.96$\pm$0.18 & $-$0.35$\pm$0.10 \\
31400$^{\rm C}$ &1.393 & 32300$\pm$\ \ 900 & 6.03$\pm$0.18 & $-$0.46$\pm$0.10 \\[1mm]
41074 &1.450 & 31100$\pm$\ \ 800 & 5.83$\pm$0.14 & $-$0.47$\pm$0.08 \\
41074$^{\rm C}$ &1.428 & 32000$\pm$\ \ 900 & 5.96$\pm$0.16 & $-$0.53$\pm$0.08 \\[1mm]
43520 &1.113 & 33700$\pm$\ \ 900 & 5.89$\pm$0.14 & $-$0.74$\pm$0.08  \\
43520$^{\rm C}$ &1.375 & 34700$\pm$1100 & 6.06$\pm$0.20 & $-$0.83$\pm$0.08 \\[1mm]
53945 & 1.277 & 34600$\pm$1600 & 5.97$\pm$0.26 & $-$0.82$\pm$0.16\\
53945$^{\rm C}$ & 1.215 & 35800$\pm$2000 & 6.10$\pm$0.30 & $-$0.81$\pm$0.16 \\[1mm]
54733 &1.021 & 34900$\pm$1300 & 6.02$\pm$0.20 & $-$0.03$\pm$0.12 \\
54733$^{\rm C}$ &1.008 & 34400$\pm$1000 & 6.00$\pm$0.20 & $-$0.16$\pm$0.12 \\[1mm]
55158 &1.937 & 32500$\pm$1000 & 5.83$\pm$0.18 & $-$0.53$\pm$0.12 \\
55158$^{\rm C}$ &2.068 & 33100$\pm$1200 & 5.88$\pm$0.24 & $-$0.69$\pm$0.14\\[1mm]
56896 &0.991 & 31900$\pm$\ \ 900 & 5.74$\pm$0.12 & $-$0.67$\pm$0.08 \\
56896$^{\rm C}$ &1.256 & 33900$\pm$\ \ 900 & 5.96$\pm$0.16 & $-$0.72$\pm$0.08\\[1mm]
59786 &1.162 & 31600$\pm$1000 & 5.91$\pm$0.18 & $-$0.60$\pm$0.10 \\
59786$^{\rm C}$ &1.244 & 32500$\pm$1000 & 6.01$\pm$0.20 & $-$0.72$\pm$0.12 \\[1mm]
75364 &1.473 & 39900$\pm$1700 & 6.32$\pm$0.24 & $-$0.39$\pm$0.16 \\
75364$^{\rm C}$ &1.316 & 39100$\pm$1700 & 6.32$\pm$0.26 & $-$0.48$\pm$0.16\\[1mm]
80690 &1.052 & 32900$\pm$\ \ 900 & 5.90$\pm$0.14 & $-$1.02$\pm$0.08 \\
80690$^{\rm C}$ &1.166 & 33300$\pm$1000 & 6.03$\pm$0.16 & $-$1.19$\pm$0.10\\[1mm]
87161 &1.112 & 33800$\pm$\ \ 800 & 5.98$\pm$0.14 & $-$0.86$\pm$0.08  \\
87161$^{\rm C}$ &1.140 & 33900$\pm$\ \ 900 & 6.11$\pm$0.18 & $-$1.00$\pm$0.10 \\[1mm]
111785 &1.421 & 40500$\pm$1000 & 5.85$\pm$0.20 & $+$0.88$\pm$0.14  \\
111785$^{\rm C}$ &1.235 & 35900$\pm$1400 & 6.28$\pm$0.22 & $+$0.68$\pm$0.10 \\[1mm]
120119 &1.515 & 33800$\pm$\ \ 900 & 5.99$\pm$0.16 & $-$0.35$\pm$0.10 \\
120119$^{\rm C}$ &1.570 & 34800$\pm$\ \ 900 & 6.07$\pm$0.18 & $-$0.45$\pm$0.10\\[1mm]
126892 &1.202 & 31500$\pm$1500 & 5.47$\pm$0.20 & $-$0.80$\pm$0.12  \\
126892$^{\rm C}$ &1.275 & 33000$\pm$1300 & 5.71$\pm$0.22 & $-$0.88$\pm$0.12\\[1mm]
181678 &1.497 & 42400$\pm$\ \ 700 & 6.27$\pm$0.18 & $+$0.18$\pm$0.12 \\
181678$^{\rm C}$ &1.298 & 42100$\pm$\ \ 700 & 6.42$\pm$0.18 & $+$0.10$\pm$0.12 \\[1mm]
189080 &1.430 & 33800$\pm$\ \ 900 & 5.93$\pm$0.16 & $-$0.44$\pm$0.10 \\
189080$^{\rm C}$ &1.416 & 34700$\pm$1000 & 6.07$\pm$0.20 & $-$0.54$\pm$0.10\\[1mm]
190398 &1.180 & 35900$\pm$1100 & 5.93$\pm$0.16 & $-$0.79$\pm$0.10  \\
190398$^{\rm C}$ &1.238 & 36700$\pm$1200 & 6.05$\pm$0.20 & $-$0.88$\pm$0.10\\
\hline
\end{tabular}
\end{table}

The helium abundances plotted in Fig.~\ref{Fig:Teff_loghe} shows
  a clear distinction between helium-poor stars with \loghe$<-1.6$
  (open squares, {\bf Group 1} hereafter) and stars with helium
  abundances close to or above solar (filled triangles, {\bf Group 2}
  hereafter).

\subsection{Spatial distribution}\label{Sec:distribution}

Moehler et al. (\cite{modr07}) noted an asymmetric spatial
distribution of the helium-rich stars. 
To verify the significance of this effect, we investigated the
spatial distribution of the faint HB stars, $B>$17, adopting the
combined Advanced Camera for Surveys (ACS) and Wide Field Imager (WFI)
photometric catalog (Castellani et al. \cite{cast07}).  We selected
candidate helium-rich and helium-poor HB stars according to the
magnitudes of the spectroscopically confirmed samples.  We assumed
helium-rich stars to be those with $B>$ 18.35 and helium-poor
stars to be those with $B\le$18.35.  The spatial distribution of the
two samples does not exhibit any significant asymmetry in the
four quadrants of the cluster.  There is mild evidence of an
overabundance of hot HB stars in general in the southeast
quadrant of $\omega$\,Cen (about 29\% versus 22--25\% in the
other quadrants), but it is within the mutual error bars. On the other
hand, the spectroscopically confirmed {\em helium-rich} EHB stars are
concentrated in the northwest quadrant of the cluster, for
which Calamida et al. (\cite{cast05}) found a lack of stars with lower
than average reddening for $u-y$ and $V-I$. We are
unable to draw a firm conclusion about the spatial distribution
of the helium-rich EHB stars because we lack sufficiently good
statistics.

\section{Evolutionary tracks}\label{sec:Tracks}
To compare the atmospheric parameters of our target
stars with the theoretical models, we computed two sets of stellar
evolutionary sequences: one with a helium-normal composition of $Y$ =
0.23 and another with a helium-rich composition of $Y$ = 0.38.  The
observed splitting of the main sequence in $\omega$\,Cen indicates
that the helium abundance in the blue main sequence (bMS) stars is
larger by $\Delta Y\approx$0.15 than the helium abundance in the red
main sequence (rMS) stars (Piotto et al. \cite{pivi05}).  Thus our
helium-normal and helium-rich sequences should represent the evolution
of the rMS and bMS stellar populations in $\omega$\,Cen, respectively.
The heavy element abundance $Z$ for each of our two helium abundances
was determined from the [Fe/H] values given by Piotto et
al. (\cite{pivi05}) and Villanova et al. (\cite{vipi07}) for the rMS
and bMS stars.  Adopting [Fe/H] = $-$1.68 for the rMS stars and
assuming an $\alpha$ element enhancement of [$\alpha$/Fe] = 0.3, we
found a scaled solar $Z$ value of 0.00064 for our helium-normal
composition.  In obtaining this $Z$ value, we used the prescription of
Salaris et al. (\cite{sach93}) to convert an $\alpha$-enhanced
composition into the equivalent scaled solar composition.  Using the
same procedure, we obtained a scaled solar $Z$ value of 0.0011 for our
helium-rich composition from the [Fe/H] value of $-$1.37 for the bMS
stars (Villanova et al. \cite{vipi07}).

     Stellar models for both of our compositions were evolved
continuously from the main sequence, up the RGB, through the helium
flash to the ZAHB, and then through the HB phase.  Mass loss was
included during the RGB evolution according to the Reimers
formulation, with the mass-loss parameter $\eta_R$ being varied
from 0 (no mass loss) up to the maximum value for which our models
evolved to the ZAHB without undergoing flash mixing.  The effective
temperatures of these canonical ZAHB models at the hot end of the EHB
were 32\,000\,K and 31,300\,K, respectively, for our helium-normal and
helium-rich compositions.  Thus the higher helium abundance of our
helium-rich composition did not increase the maximum effective
temperature along the canonical ZAHB.  We next computed additional
sequences with higher mass-loss rates to determine the range
in $\eta_R$ over which flash mixing occurs.  ZAHB models for the
minimum, average, and maximum values of $\eta_R$ leading to flash
mixing were constructed assuming that all of the envelope hydrogen was
burned during the mixing phase and that the envelope carbon abundance
was increased to 0.04 by mass.  These assumptions are consistent with
the flash-mixing calculations of Cassisi et al. (\cite{casc03}) and
Miller Bertolami et al. (\cite{mial08}).  The average effective
temperatures of the flash-mixed ZAHB models were 37,500\,K and
35,900\,K, respectively, for our helium-normal and helium-rich
compositions.  Thus both of these compositions predict a gap of
$\approx$5\,000\,K between the flash-mixed and the hottest canonical
ZAHB models in good agreement with the earlier results of Brown et
al. (\cite{brsw01}).  These flash-mixed ZAHB models were then evolved
through the HB phase.  Sequences with even higher mass-loss rates
failed to ignite helium and thus died as helium white dwarfs.

\begin{figure}[h!]
\includegraphics [height=1.0\columnwidth,angle=270]{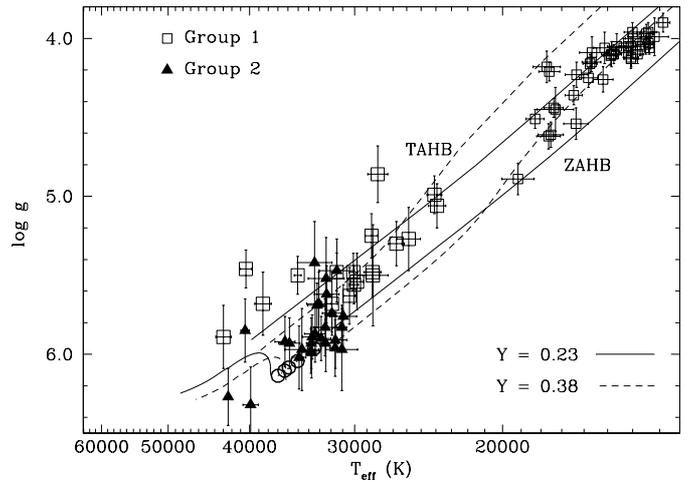}
\caption[]{The effective temperatures and surface gravities derived
  for our target stars (formal errors multiplied by 2, see text for
  details).  Helium-poor and helium-rich stars are marked by open
  squares and filled triangles, respectively. The stars with super-solar helium are
  shown with the parameters derived from models without C/N
  enhancement.  The solid lines mark the canonical HB locus ($Y$ =
  0.23) and the dashed lines mark the helium-enriched HB locus ($Y$ =
  0.38, see text for details).  The tracks for a late hot flasher
  (same line types as for the ZAHB) show the evolution of these stars
  from the zero-age HB (ZAHB) towards helium exhaustion in the core
  (terminal-age HB = TAHB). The dotted line connects the series of
  ZAHB models computed by adding a hydrogen-rich layer to the surface
  of the canonical ZAHB model of the late hot flasher.  The open circles
  mark -- with decreasing temperature -- hydrogen layer masses of $0,
  10^{-7}, 10^{-6}, 10^{-5}$, and $10^{-4}$\Msolar\ (for details see Moehler
  et al. \cite{mosw02}).  }\label{Fig:Teff_logg}
\end{figure}

\section{Results and discussion}\label{sec:Results}
\subsection{Helium-poor stars (Group 1)\label{sec:he-poor}}
\begin{figure}[h!]
\includegraphics [height=1.0\columnwidth,angle=270]{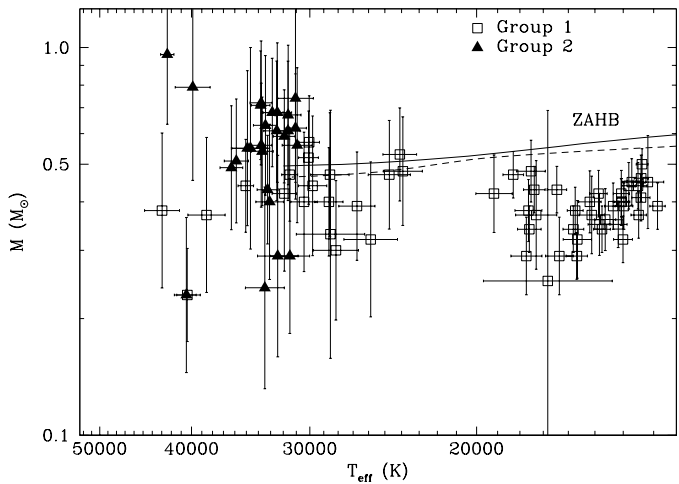}
\caption[]{The effective temperatures and masses derived for our
target stars (formal errors multiplied by 2, see text for details).
Helium-poor and helium-rich stars are marked by open squares and
filled triangles, respectively. The stars with super-solar helium 
abundances are shown with the parameters derived from models without
C/N enhancement.  The lines mark the zero-age horizontal branch for
$Y$ = 0.23 (solid) and 0.38 (dashed, see text for
details).}\label{Fig:Teff_logM}
\end{figure}

The helium-poor stars plotted in Figs.~\ref{Fig:Teff_logg} and
\ref{Fig:Teff_logM} exhibit the same behaviour as hot HB stars
and EHB stars in other globular clusters (see Moni Bidin et
al. \cite{momo07} for a recent discussion). While comparing both
effective temperature and surface gravity with the tracks 
implies helium enrichment, the too-low masses clearly indicate that
the results are not trustworthy. As mentioned in other papers, we
suspect that the diffusion in the stars' atmospheres creates abundance
ratios that are not correctly described by the model atmospheres
we use.

It is noticeable, however, that all helium-poor stars with
  effective temperatures above 32\,000\,K (the end of the canonical
  ZAHB) have evolved away from the horizontal branch
  (cf. Fig.~\ref{Fig:Teff_logg}).
\subsection{Helium-rich stars (Group 2)\label{sec:he-rich}}
The helium-rich stars cover the temperature range between the hot end
of the ZAHB and the late hot flasher region. As already discussed by
Moehler et al. (\cite{modr07}), diffusion acting in a late hot flasher
would move any remaining hydrogen to the surface, while at the same time
reducing the effective temperature. This behaviour is consistent with
that observed in Figs.~\ref{Fig:Teff_loghe} and \ref{Fig:Teff_logg}.
\begin{figure}[h!]
\includegraphics [width=1.0\columnwidth,angle=0]{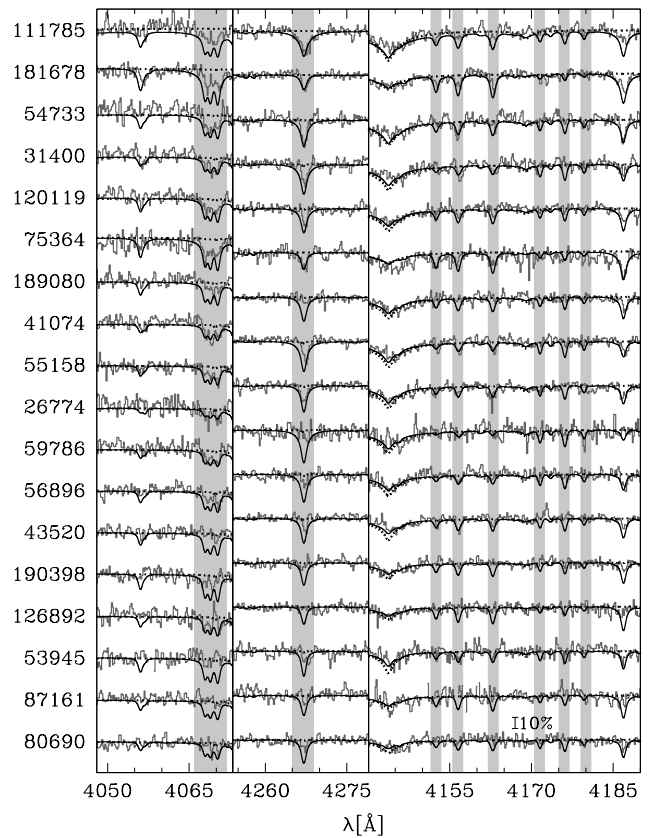}
\caption[]{Spectra of all stars with super-solar helium abundances in
 spectral regions where strong carbon lines (marked with grey shading)
 are expected. The helium abundance decreases from top to bottom and
 the stars' names are given.
The black solid line marks the model spectrum with a carbon/nitrogen
 abundance of 3\%/1\% by mass, while the dotted line (horizontal
 except for the \ion{He}{i} line at 4144\AA) indicates the model
 spectrum with the cluster carbon and nitrogen abundances. We always
 plot the model spectra that best fit the observed helium and
 hydrogen lines.}\label{Fig:carbon}
\end{figure}

In Fig.~\ref{Fig:carbon}, we show the spectra of the stars with
super-solar helium abundance where the helium abundance 
decreases from top to bottom. We overplot model spectra with
the cluster carbon and nitrogen abundance (dotted lines, mostly
just horizontal) and with a carbon/nitrogen abundance of 3\%/1\% by
mass, respectively (solid black lines). The model spectra were derived
by fitting the helium and hydrogen lines of the spectra, {\em not} the
carbon lines. Obviously the most helium-rich stars show a strong
tendency towards a high carbon abundance, which can so far only be
explained by the late hot flasher scenario. This would also explain
the rather abrupt change in helium abundance at effective temperatures
hotter than the hot end of the canonical HB.

\section{Conclusions}\label{sec:conclusions}
From our spectroscopic analysis of the spectra of hot horizontal
branch stars, we derive the following conclusions:
\begin{enumerate}
\item We have found no evidence of close binaries among
  our targets.
\item The effective temperatures and surface gravities of the
  helium-poor HB stars below 20\,000\,K are at first glance 
  indicative of helium enrichment. The too-low masses derived from these
  parameters, however, render these results dubious. This, however,
  does {\em not} rule out the presence of helium-enriched stars in
  this temperature range. 
\item The parameters of the stars in Group~2 agree well with the
  predictions of the late hot flasher scenario, if one allows for some
  residual hydrogen and diffusion effects. Strong arguments in favour
  of this scenario are the presence of stars with helium abundances in
  excess of the predictions of D'Antona et al. (\cite{daca10})
  and
   clear evidence of carbon enrichment by at least a factor of 10 in
  the more helium-rich stars. Additional support is provided by
  the evolved status of
  all helium-poor stars above 32\,000\,K (the hot end of the
  canonical ZAHB). This does not rule out the possibility
  that the blue hook stars belong to the helium-enriched
  sub-population, but this helium enrichment alone cannot explain the
  observed parameters of the stars (as already stated by Moehler et
  al. \cite{modr07}).
\end{enumerate}

\begin{acknowledgements}We thank the staff at the Paranal observatory and at
  ESO Garching for their excellent work, which made this paper
  possible. Two of us (GB, AC) were partially supported by Monte dei
  Paschi di Siena (P.I.: S. Degl'Innocenti) and by PRIN-MIUR2007
  (P.I.: G. Piotto). This research has made use of NASA's Astrophysics
  Data System. We thank the anonymous referee for helpful comments.
\end{acknowledgements}

\end{document}